\documentclass[doublecol]{epl2} 
\title{A network-specific approach to percolation in complex networks with bidirectional links}
\author{D. Taylor\inst{1} \and J. G. Restrepo\inst{1} }
\shortauthor{D. Taylor and J. G. Restrepo}
\institute{                    
  \inst{1} Department of Applied Mathematics, University of Colorado, Boulder, Colorado 80309, USA\\
}
\pacs{64.60.ah}{Percolation}
\pacs{89.75.-k}{Complex systems}
\pacs{05.70.Jk}{Critical point phenomena}

\abstract{
Methods for determining the percolation threshold usually study the behavior of network ensembles and are often restricted to a particular type of probabilistic node/link removal strategy.
We propose a network-specific method to determine the connectivity of nodes below the percolation threshold and offer an estimate to the percolation threshold in networks with bidirectional links.
Our analysis does not require the assumption that a network belongs to a specific ensemble and can at the same time easily handle arbitrary removal strategies (previously an open problem for undirected networks).
In validating our analysis, we find that it predicts the effects of many known complex structures ({\it e.g.}, degree correlations) and may be used to study both probabilistic and deterministic attacks.
}

\begin{document}
\maketitle

\section{Introduction} 

The study of percolation in complex networks has broad applications including epidemic spreading \cite{outbreak}, propagation of excitation in neural networks \cite{dans}, and robustness of networks to random failure \cite{Yeh} or strategic attack \cite{bias,DI}. A central problem is estimating the {\it percolation threshold}, the critical fraction of nodes or links of an initially connected network that must be removed to disintegrate it into small disconnected fragments. Knowledge of how a network fragments can improve strategies for designing attack \cite{bias,DI} and immunization techniques \cite{immune} or increasing network robustness \cite{dane,Malicious}. 

Several studies have proposed techniques to estimate the percolation threshold of a network for various situations \cite{bias,Cohen00,Newman01,Boguna05,wPerc, MarkUndir2}. These studies typically use {\it ensemble approaches}, where one studies the typical behavior of a set of networks satisfying some set of properties. Common ensembles include: (i) networks with fixed degree sequence generated from, for example, the configuration model \cite{config}; (ii) networks with an expected degree sequence generated with, for example, the Chung-Lu model \cite{ChungLu}; and (iii) {\it Markovian} networks with correlations between nearest neighbors, where correlations may be captured by the probability $P({\bf  d}'| {\bf d})$ that a node with degree ${\bf d}$ is connected to a node of degree ${\bf  d}'$. Here ${\bf d} = (d^{in},d^{out})$ denotes the the number of incoming ($d^{in}$) and outgoing ($d^{out}$) links at a given node.  While significant progress has been made in the study of such ensembles \cite{Newman01,Cohen00,MarkUndir2,Boguna05, bias}, we note two important limitations of the ensemble approach. 
(i) Given a single network, it is not clear what ensemble should be selected to capture the properties of the network. Typically, networks found in applications contain various structural properties ({\it e.g.}, correlations \cite{MarkUndir2}, clustering \cite{clustering}, and community structure \cite{modularity}) that are not always accounted for in the ensembles. A related problem is that it has been recently observed \cite{ensemble} that, given an ensemble of networks, some network properties can vary significantly within the ensemble. Thus, it is not clear that ensemble approaches give the best description of a single network found in practice. 
(ii) Some ensemble theories are impractical when applied to individual real networks. For example, theories to estimate the percolation threshold in Markovian ensembles \cite{Boguna05} require the estimation of a potentially dense and very large, $(d^{in}_{max}d^{out}_{max}) \times (d^{in}_{max}d^{out}_{max})$ matrix $P({\bf  d}| {\bf d}')$, which is difficult to obtain from a single network.

In this Letter, we develop a network-specific theory for percolation in networks with bidirectional links. In contrast to previous approaches to network-specific estimates of the percolation threshold which were restricted to purely directed networks \cite{wPerc}, our work has a broader range of applicability as it can be applied to undirected networks and directed networks with bidirectional links.
Our method is based on an analysis of the network's adjacency matrix, which is often known or can be estimated in important applications ({\it e.g.}, the power grid \cite{power} and air transportation networks \cite{airports}).
Besides relaxing the ensemble assumptions of previous research ({\it e.g.}, that the network is strictly Markovian), one significant advantage of this approach is that it can easily account for arbitrary strategies of node/link removal. Network-specific approaches are therefore well suited for developing network-specific attack/defense strategies, immunization techniques, etc.
In addition to estimating the percolation threshold, we predict the expected number of nodes accessible to each node after the network disintegrates. This has various applications such as predicting the outbreak size of an epidemic \cite{outbreak}.
We finally show that our method may be used to study the fragmentation of a network subject to either probabilistic or deterministic attack. 

\section{Analysis}

We formalize weighted percolation ({\it i.e.,} in which nodes and/or links are retained with arbitrary probabilities) as follows: for a network with $N$ nodes described by a possibly asymmetric adjacency matrix $A$ ($A_{nm}=1$ if a link exists from node $n$ to node $m$ and $A_{nm} = 0$ otherwise), node $n$ is retained with probability $q_n$, and the directed link from node $n$ to node $m$ is retained with probability $p_{nm}$. Letting $q=N^{-1}\sum_n q_n$ denote the {\it average node retention probability}, $qN$ nodes are expected to remain after a realization of this process (referred to as a {\it percolation trial}). 
Unweighted node percolation corresponds to $q_n=q < 1$ and $p_{nm}=1$, while
unweighted link percolation corresponds to $q_n=1$ and $p_{nm}=p<1$.
For our analysis, it is useful to introduce a matrix $\hat A$ with entries defined by $q\hat{A}_{nm} = q_m p_{nm} A_{nm}$, which represents the probability that a link exists from node $n$ to node $m$, given that node $n$ is retained. 
Because our analysis depends only on the matrix $\hat A$, it is applicable to link, node, and mixed ({\it i.e.,} simultaneous link and node) percolation. However, for the remainder of this Letter we consider only node percolation, $p_{nm}=1$.

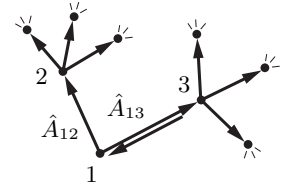
\begin{figure}[h]
\begin{picture}(2000,70)
    \put(-00,0){\onefigure[width=.4\linewidth]{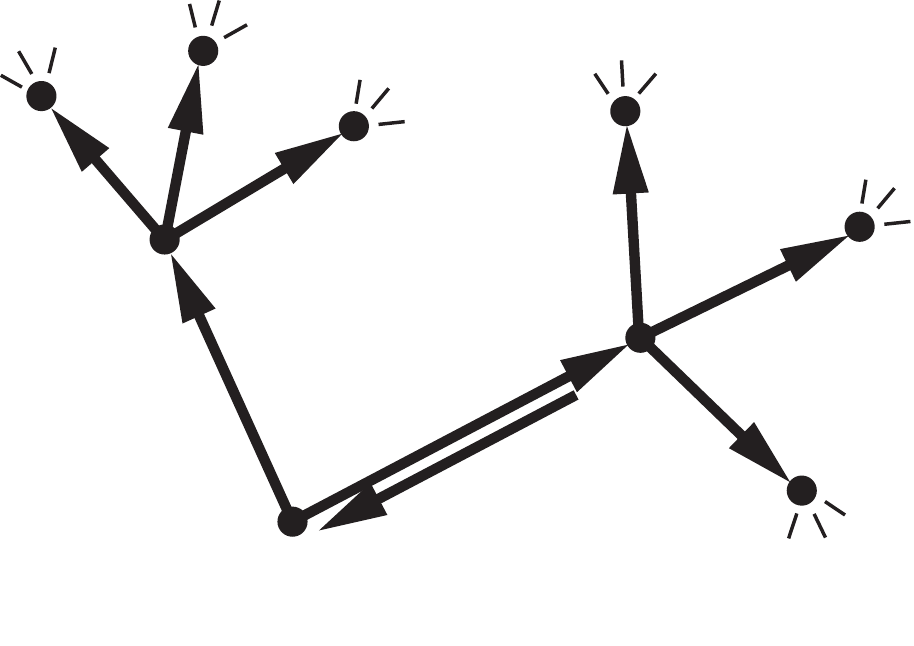}}
     \small{  \put(102,3){$1$}
    \put(83,40){$2$}
    \put(137,37){$3$}
    \put(85,20){$\hat{A}_{12}$}
    \put(110,28){$\hat{A}_{13}$}}
  \end{picture}
  \setlength{\abovecaptionskip}{-0.5cm}
\setlength{\belowcaptionskip}{-0.1cm}
\caption{When computing $ s^{out}_1$, to compensate for the over-counting of nodes due to the bidirectional link $1\leftrightarrow 3$, $\beta_{13}$ reduces the contribution of $ s^{out}_3$ on $ s^{out}_1$. We approximately have $\beta_{13}\sim3/4$ and $ s^{out}_1=1+q\hat{A}_{12} s^{out}_2+q\hat{A}_{13}\beta_{13} s^{out}_3$.}
\label{g_n_cartoon}
\end{figure}

For a given node-targeting strategy, defined as a set of retention probabilities $\{q_n\}$, we are interested in the size $s$ of the {\it largest strongly-connected component} (LSCC), the largest subset of nodes so that any node in the subset is reachable from any other node in the subset. The percolation threshold $q^*$ is defined as the value of $q$ such that $s\ll N$ for $q < q^*$ (the {\it subcritical regime}) and $s\sim N$ for $q > q^*$ (the {\it supercritical regime}). For a particular removal strategy, the subcritical regime may be analyzed by noting that after a percolation trial, only a fraction of the network is reachable from a given node $n$ following directed links. Following \cite{Newman01} we define, for a given percolation trial, the {\it out-component} of node $n$ as the set of nodes that may be reached from node $n$ via the remaining network (including node $n$) and define $s^{out}_n$ as the size of the out-component of node $n$ averaged over many percolation trials.
To motivate subsequent analysis, consider first the case when the network is a directed tree. In that case $s^{out}_n$ satisfies the relation $s^{out}_n = 1 + \sum_m q\hat A_{nm} s^{out}_m$,
where the right hand side counts the nodes reachable from node $n$ by counting the nodes reachable from its neighbors, and adds $1$ to account for node $n$ itself. The same expression approximately applies to directed networks that are locally tree-like \cite{gleesonmucha} and leads to the results in ref.~\cite{wPerc}, $q^*\approx \hat\lambda^{-1},$ where $\hat\lambda$ is the principal, or Perron-Frobenius, eigenvalue of $\hat{A}$ (i.e., $\hat A{\bf\hat u} =\hat\lambda{\bf\hat u}$). When links are allowed to be bidirectional, however, the expression above overestimates the size of $s^{out}_n$, since the terms $s^{out}_m$ on the right hand side might include nodes that are reachable by following links back into node $n$ (see fig.~\ref{g_n_cartoon}). 

To correct for this over-counting of nodes, we heuristically modify the contribution of $s^{out}_m$ on the right hand side by a factor $\beta_{nm}$ (to be determined),
\begin{equation}
\setlength{\belowdisplayskip}{0.2cm}
\setlength{\abovedisplayskip}{0.2cm}
s^{out}_n = 1 + \sum_m q\hat A_{nm} \beta_{nm}s^{out}_m.\label{g1}
\end{equation}
To determine a self-consistent expression for $\beta_{nm}$, we note that from eq.~(\ref{g1}) the relative contribution of the out-component of node $m$ on $s^{out}_n$ is ${q\hat{A}_{nm}\beta_{nm}s^{out}_m }/{s^{out}_n}$. Therefore, to reduce the contribution of $s^{out}_m$ on $s^{out}_n$ to account for the branch returning to node $n$ (if present), we let $\beta_{nm} =1-{q\hat{A}_{mn}\beta_{mn}s^{out}_n}/ {s^{out}_m}$. Inserting here the corresponding expression for $\beta_{mn}$ and solving for $\beta_{nm}$ we obtain
\begin{equation}
\setlength{\belowdisplayskip}{0.2cm}
\setlength{\abovedisplayskip}{0.2cm}
	 \beta_{nm} =(1-q^2\hat{A}_{nm}\hat{A}_{mn})^{-1}\left(1-\frac{q\hat{A}_{mn}s^{out}_n}{s^{out}_m}\right). \label{beta1}
 \end{equation}
After substitution of eq.~(\ref{beta1}) into eq.~(\ref{g1}), we find
\begin{equation}
	 {\bf s}^{out} =[I-D(q)]^{-1}{\bf y} \label{g2},
 \end{equation}
where ${\bf s}^{out}=[s^{out}_1,...,s^{out}_N]^T$, $I$ is an identity matrix of size $N$, ${\bf y}$ is a vector with entries
\begin{equation}
\setlength{\belowdisplayskip}{0.2cm}
\setlength{\abovedisplayskip}{0.2cm}
y_n = \Big[1+q^2\sum_k \hat{A}_{nk}\hat{A}_{kn}(1-q^2\hat{A}_{nk}\hat{A}_{kn})^{-1}\Big]^{-1},
\end{equation}
and $D(q)$ is a matrix with entries
\begin{equation}
\setlength{\belowdisplayskip}{0.2cm}
\setlength{\abovedisplayskip}{0.2cm}
     D_{nm}(q)=q\hat{A}_{nm}y_n(1-q^2\hat{A}_{nm}\hat{A}_{mn})^{-1} .\nonumber 
 \end{equation}
Given a removal strategy, eq.~(\ref{g2}) can be solved to obtain the expected out-component size for each node. To obtain an estimate for the percolation threshold, note that eq.~(\ref{g2}) requires the invertibility of the matrix $I-D(q)$. This matrix is invertible when $\lambda_{D(q)}<1$, where $\lambda_{D(q)}$ is the principal eigenvalue of $D(q)$. As $\lambda_{D(q)}\to1^-$ the out-component sizes diverge as $s^{out}_n\sim [1-\lambda_{D(q)}]^{-1}  w_n$, where ${\bf  w} $ is the principal eigenvector of $D(q)$. A similar argument can be made for the divergence of the in-component sizes. Since the LSCC above the percolation threshold can be thought of as the set of vertices with infinite in- and out-components \cite{Boguna05}, we predict the percolation threshold as
\begin{equation}
\setlength{\belowdisplayskip}{0.2cm}
\setlength{\abovedisplayskip}{0.2cm}
 q_D^*=\min_{q\in[0,1]}\{q:\lambda_{D(q)}=1\}.\label{q1}
 \end{equation}
We note that if there are no bidirectional links, $\hat A_{nm}\hat A_{mn} = 0$ and $D(q) = q\hat A$, and the results of ref.~\cite{wPerc} are recovered. 
While one may solve eq.~(\ref{q1}) numerically,
it is both practical and insightful to approximate eqs.~(\ref{beta1}-\ref{q1}) for large $s^{out}_n$ and small $q$. 
%
Letting $s^{out}_n\gg1$ and $\beta_{nm}\sim1$ in eq.~(\ref{g1}) yields the approximate eigenvalue problem $s^{out}_n\approx q\sum_m\hat{A}_{nm}s^{out}_m$. 
It follows that ${\bf s}^{out}\propto\hat{{\bf u}}$. Upon substitution we find $q\sim\hat{\lambda}^{-1}$ under these conditions, yielding to first order $\beta_{nm} \approx 1 -\hat{\lambda}^{-1}\hat{A}_{mn}\hat{u}_n/ \hat{u}_m $. Defining 
\setlength{\belowdisplayskip}{0.2cm}
\setlength{\abovedisplayskip}{0.2cm}
\begin{equation}
C_{nm}=\hat{A}_{nm}\left(1-\frac{\hat{A}_{mn}\hat{u}_n}{\hat{\lambda}\hat{u}_m} \right), \label{C}
\end{equation}
with principal eigenvalue equation $C{\bf z} = \lambda_C \bf z$ and using ${\bf y}\approx {\bf 1} = [1,1,\dots,1]^T$, we obtain the predictions
\setlength{\belowdisplayskip}{0.2cm}
\setlength{\abovedisplayskip}{0.2cm}
\begin{eqnarray}
      {\bf s}^{out}& \approx&(I-qC)^{-1}{\bf 1} , \label{g3}\\
       q_C^*&\approx&\lambda_C^{-1}\label{q2} .
\end{eqnarray} 
In addition to offering simplified predictions for $s^{out}_n$ and $q^*$, for unweighted percolation ({\it i.e.,} $\hat A=A$ and $\bf \hat u =\bf u$) these estimates allow us to bound $\lambda_C$ using the principal eigenvalue $\lambda$ of the network adjacency matrix $A$ ({\it e.g.,} $A\bf u=\lambda \bf u$). Direct application of the Bauer-Fike Theorem \cite{Golub} for the limiting case of an undirected network yields $|\lambda_C-\lambda|\le ||\lambda^{-1}UAU^{-1} ||_2=1$, where $U=\mbox{diag}[ u_1,\dots, u_N]$. Finally, considering ${\bf1}^T$$C$$\bf{u}$ and using $\bf z \sim {\bf u}$ yields $\lambda_C\approx \lambda-1$.
One implication of these results is that $q^*\to0$ for large $\lambda$, which is consistent with the lack of an unweighted percolation threshold for well-connected networks such as scale-free networks \cite{Cohen00}.
%
%
{We note that eqs.~(\ref{g3}) and (\ref{q2}) are in best agreement with eqs.~(\ref{g2}) and (\ref{q1}) near $q=q^*$ and when the network is strictly undirected or strictly directed.}

\begin{figure}[t] 
\onefigure[width=1\linewidth]{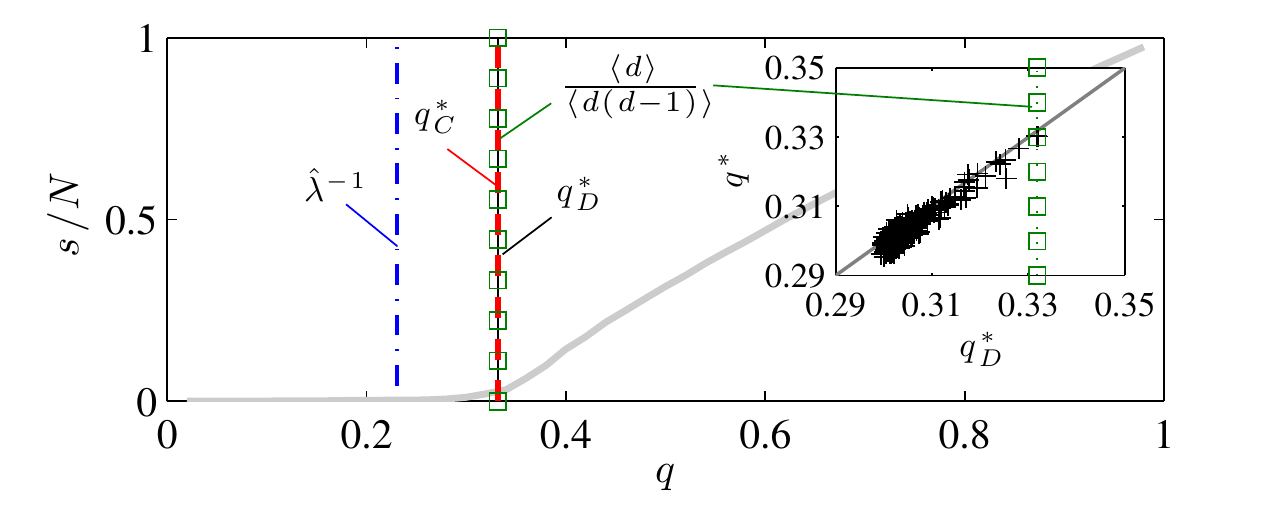}
\setlength{\abovecaptionskip}{-.0cm}
\setlength{\belowcaptionskip}{-.3cm}
\caption{(Colour online) The relative size of the giant component, $s/N$ (solid grey line), is shown for unweighted percolation on an ER network \cite{ER} (see text). {Our predictions to $q^*$ given by eq. (\ref{q1}) and eq. (\ref{q2}) are given by vertical solid black and dashed red lines }and are in good agreement with the undirected ensemble result \cite{Newman01} (green squares). The network-specific prediction for directed networks, $\hat\lambda^{-1}$\cite{wPerc} (blue dot-dashed line), is shown to be inaccurate. The inset shows experimental values for $q^*$ 
vs. eq.~(\ref{q1}) for networks within an ensemble. Unlike the ensemble approach, variation in $q^*$ is naturally accounted for by our network-specific approach.}
\label{11}
\end{figure}

\section{Examples}
{In what follows, we will motivate the need for our theory, recover previous results, explore several applications, and illustrate the robustness of our analysis to complex structures in networks. We will consider both computer-generated and real-world networks.}

{We first highlight the need for a network-specific method for undirected networks and show that unlike the ensemble approach, a network-specific method captures variability in $q^*$ across an ensemble. }We consider percolation in an uncorrelated, random network formed by retaining the giant component from an Erd\"os-R\'enyi process \cite{ER} (henceforth referred to as an ER network) with $N=10^4$ nodes and $3N$ links. 
In fig.~\ref{11} we show the fraction of retained nodes in the LSCC, $s/N$, as a function of $q$ (solid grey curve). Our predictions for $q^*$ given by {eq. (\ref{q1}) (black line) and eq. (\ref{q2}) (red dashed line)} and the undirected ensemble theory \cite{Newman01} (green squares)  work well, whereas the network-specific theory for directed networks, $\hat\lambda^{-1}$ \cite{wPerc} (blue dot-dashed line), does not, as expected. 
The inset shows experimental values for $q^*$ versus our prediction using eq.~(\ref{q1}) for an ensemble of uncorrelated networks obtained by rewiring the ER network while retaining a fixed degree sequence, $[d_1,d_2,\dots, d_N]$. The rewiring process is similar to that in refs.~\cite{MarkUndir2,DI}, except an additional step is taken to ensure the resulting network contains all nodes in its LSCC.
Note that whereas our network-specific method naturally accounts for variability in $q^*$ across the ensemble, the ensemble approach (squares) cannot, predicting $q^*\approx0.33$ for all members of the ensemble. 

{Experimental values for $q^*$ in this inset and remaining figures were found using extrapolation over 6 overlapping intervals of width 0.03 that span $[0.03,0.13]$ to find the intercept with the horizontal axis. This yielded a mean and standard deviation (shown when significant). While the actual percolation threshold $q^*$ is often well defined for model networks in the $N \to \infty$ limit (e.g., by using traditional finite-size scaling arguments), one can only estimate it for real-world networks which have a fixed and finite size. }
We also note that while $q^*$ can often be extracted by examining the size of the second-largest cluster, this approach was observed to significantly overestimate $q^*$ for the relatively small networks considered here for which $N<10^5$. 

\begin{figure}[t]
\setlength{\belowcaptionskip}{-0.2cm}
\setlength{\abovecaptionskip}{-0.2cm}
\onefigure[width=\linewidth]{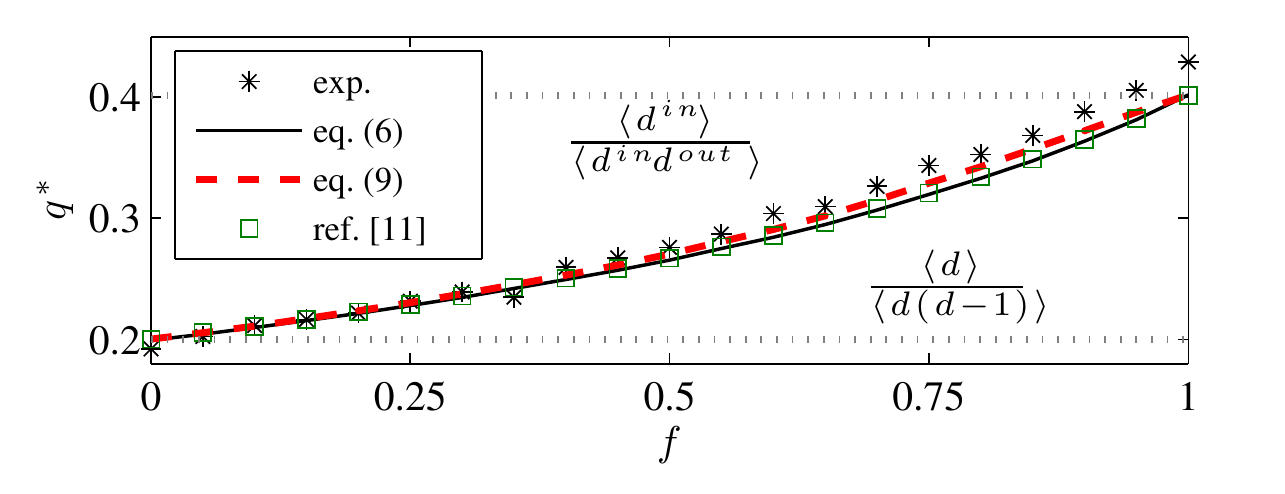}
\caption{(Colour online) Equations~(\ref{q1}) and eq.~(\ref{q2}) agree well with ref. \cite{Boguna05} and experimental values for predicting the unweighted percolation threshold $q^*$ of an uncorrelated random network. $f$ denotes the fraction of links that are directed, which increases from 0 (undirected) to 1 (directed). }
\label{3}
\end{figure}

{We next illustrate our analysis in networks with bidirectional links and recover previous results for unweighted percolation in uncorrelated networks. }Letting $f$ denote the fraction of directed links, we begin with an undirected ({\it i.e.,} $f=0$) ER network with $N=10^4$ nodes and 5N links and iteratively replace randomly chosen undirected links with directed links in a random orientation until the network is strictly directed ({\it i.e.,} $f=1$). One can observe in fig.~\ref{3} that our predictions by eqs.~(\ref{q1}) and (\ref{q2}) agree well with the ensemble result \cite{Boguna05} and experimental values. {Note that the predictions of eq. (\ref{q2}) agree with those of eq. (\ref{q1}) very well for $f=0$ or 1 and that maximal disagreement occurs near $f\sim0.7$. 
We note that one disadvantage of extrapolating on a fixed range ({\it e.g.,} $s/N\in[0.03,0.13]$) while varying $q^*$ is that a drift occurs due to the combination of finite-size affects and the fact that the slope of an $s(q)$ curve just above $q^*$ varies monotonically with $q^*$. This is one cause for the disagreement between theory and experiment in fig. \ref{3} for large $f$.}

We now recover the ensemble predictions for strictly undirected and strictly directed networks shown by horizontal lines in fig.~\ref{3}. For $f=0$, our result $\lambda_C\approx\lambda-1$ and the undirected mean-field (MF) result $\lambda\approx \langle d^2\rangle /\langle d \rangle$ lead to $q^*\approx \langle d \rangle/\langle d(d-1) \rangle $ \cite{Cohen00}. For $f=1$ we have $A_{nm}A_{mn}=0$ and $C= A$, which recovers $q^*\approx \lambda^{-1}$ \cite{wPerc}. Again, the MF result $\lambda^{-1}\approx\langle d^{in}\rangle / \langle d^{in}d^{out} \rangle  $ \cite{wPerc} recovers the result of ref.~\cite{Newman01}. 

\begin{figure}[t] 
\setlength{\abovecaptionskip}{-0.1cm}
\setlength{\belowcaptionskip}{-0.2cm}
\onefigure[width=\linewidth]{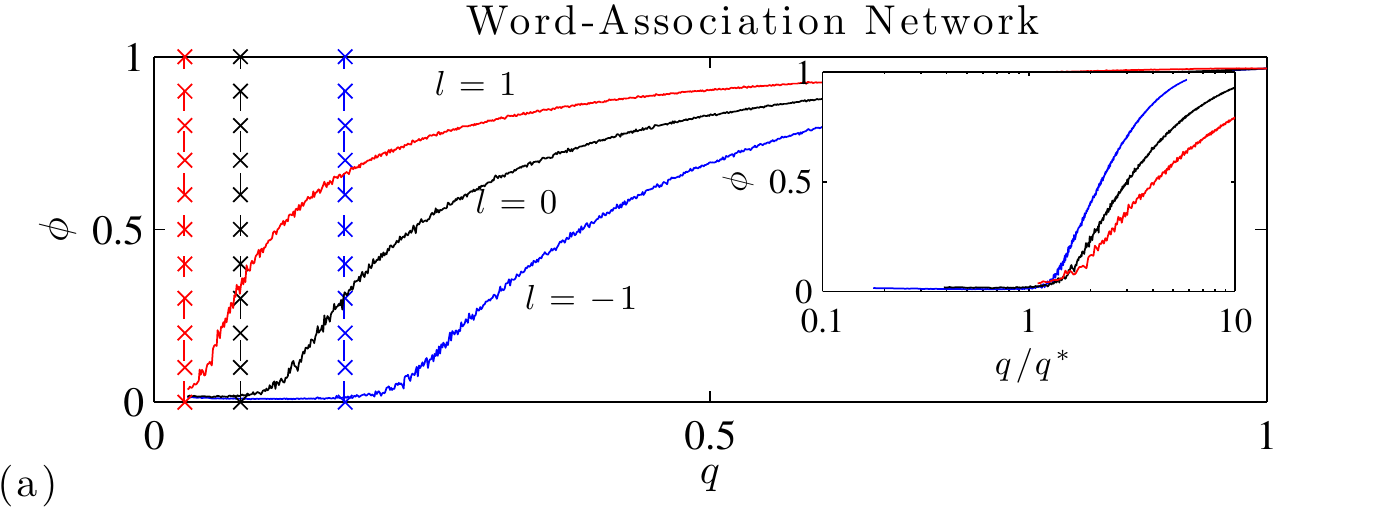}
\onefigure[width=\linewidth]{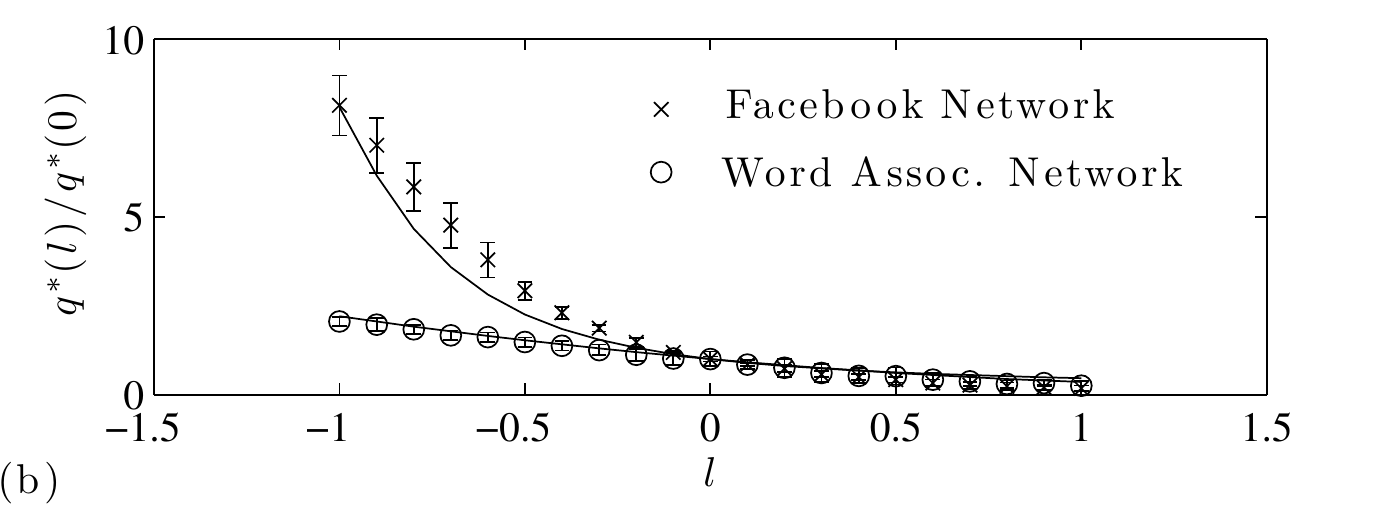}
\caption{(Colour online) (a) Average values for $\phi=s/(qN)$ over 32 trials are shown for weighted percolation with $q_n\propto d_n^l$ on the scale-free WA network \cite{Word}. {Vertical dashed lines and x's represent the predictions of eq.~(\ref{q1}) and eq.~(\ref{q2}). }The inset shows the same quantities with the $q$-axis normalized by eq.~(\ref{q1}) for each $l$ value.
(b) The dependency of $q^*$ on $l$ is shown for the WA (circles) and FB networks \cite{FB} (x's).
}
\label{5}
\end{figure}

Our next examples explore targeted attacks, where we let $q_n\propto d_n^l$ for $l\in\{-1,0,1\}$. Here $d_n=(d_n^{in}+d_n^{out})/2$ denotes the degree of node $n$ averaged over incoming and outgoing links. Preferentially removing nodes with large degrees ({\it i.e.,} $l<0$) can model probabilistic attacks \cite{bias} and biased infrastructure failure \cite{Yeh}, whereas preferentially removing nodes with small degrees ($l>0$) may represent a non-obtrusive degradation. We consider real-world networks for which previous methods are either impractical or not justified for the reasons mentioned in the introduction. In fig.~\ref{5}a we show the relative size of the LSCC, $\phi=s/(qN)$, for weighted percolation on a directed Word-Association (WA) Network \cite{Word}. {Vertical dashed lines and x's represent the predictions of eq.~(\ref{q1}) and eq.~(\ref{q2}).} The inset shows this graph with the horizontal axis respectively normalized by the value of $q^*$ found from eq.~(\ref{q1}) for each targeting strategy, where a LSCC of size $s \sim N$ appears at $q/q^*\sim1$ for all curves. In fig.~\ref{5}b, we show the prediction by eq.~(\ref{q1}) (solid lines) and observed (symbols) values of $q^*$ as a function of $l$, normalized by the value of $q^*$ at $l=0$, for the WA network \cite{Word} and an undirected  network of Facebook (FB) friendships at Caltech \cite{FB}. Being a well-connected network with power-law degree distribution, the FB network has a very small unweighted percolation threshold [$q^*(0)\sim0.01$, not shown]. However, preferentially removing nodes with large degree can yield nontrivial thresholds with nearly a ten-fold increase [$q^*(-1)\sim0.1$] .

\begin{figure}[h] 
\setlength{\belowcaptionskip}{-0.3cm}
\setlength{\abovecaptionskip}{0.0cm}
\onefigure[width=.9\linewidth]{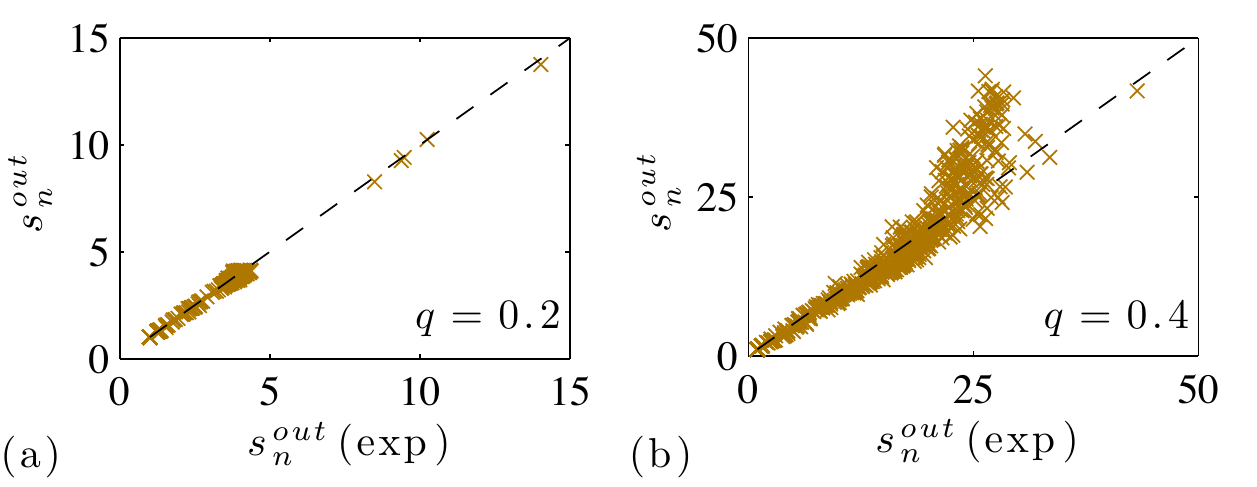}
\caption{(Colour online) (a) For weighted percolation on a P2P network \cite{Gnutella} with retention probability $q_n\propto d_n^{-0.5}$ and average retention rate $q=0.2$, the values of $s_n^{out}$ predicted by eq.~(\ref{g2}) (x's) agree well with experimental values, $s^{out}_n$ (exp), which were averaged over $2^{16}$ percolation trials.
(b) Similar results for $q=0.4$, where some deviations appear (see text).}
\label{2}
\end{figure}

Our next example illustrates eqs.~(\ref{g2}) and (\ref{g3}) in a peer-to-peer (P2P) network of file downloads \cite{Gnutella} with $N = 6301$ nodes for weighted site-percolation with $q_n\propto d_n^{-0.5}$. In this example $s_n^{out}$ can be interpreted as the expected number of nodes infected by a computer virus released by user $n$ and preferentially targeting nodes with large degree \cite{outbreak}. In fig.~\ref{2} we show a sample of the values of ${s}^{out}_n$ predicted by eq.~(\ref{g2}) (x's) [which agrees with eq.~(\ref{g3})] versus their experimental values, $s^{out}_n$ (exp). The prediction given by eq.~(\ref{g2}) (x's) is very accurate for the average removal probability $q=0.2$ shown in fig.~\ref{2}a. For the larger value $q=0.4$, our prediction deviates somewhat from the observed values as shown in fig.~\ref{2}b. This is expected because $s^{out}_n$ is predicted to diverge at $q^*\sim0.476$, but experimental values are bounded by the finite network size $N$, so the predicted value must become larger than the observed value as $q\to q^*$. 

\begin{figure}[t] 
\onefigure[width=1\linewidth]{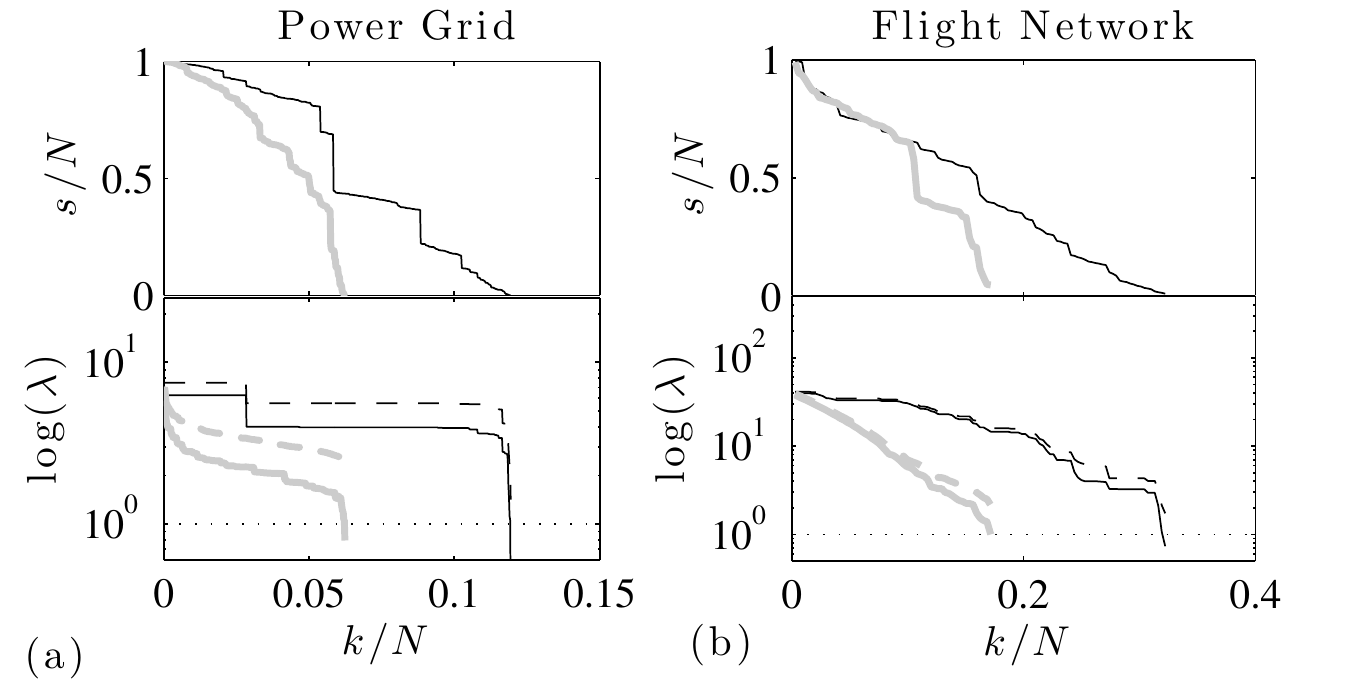}
\caption{Malicious attacks were simulated on two significant undirected networks: (a) a power grid for the U.S. \cite{power} and (b) a flight network for US Airlines \cite{airports}, where the node with largest dynamical importance \cite{DI} was iteratively removed (along with nodes disconnected from the LSCC after its removal). The relative size of the LSCC is shown in the top panels (grey lines) as a function of the fraction of removed nodes, $k/N$. The relevant eigenvalues for the networks' adjacency matrices, {\it i.e.,} $\lambda$ (dashed grey lines) and $\lambda_C$ (solid grey lines) are shown in the bottom panel as a function of the fraction $k/N$ of removed nodes. Black lines indicate the same variables under random node removal. The disappearance of the LSCC corresponds to $\lambda_C\approx1$ (horizontal dotted lines). }
\label{Attack}
\end{figure}

{In our next example we apply our results to network fragmentation under deterministic attack. }We consider two important undirected networks,  a power grid \cite{power} and an airline transit network \cite{airports}, and point out that knowledge of how they fragment is essential for taking measures of protection.
Simulated attacks were implemented by iteratively removing the node $n$ corresponding to the largest dynamical importance, $\mbox{DI}_n\propto u_nv_n$ \cite{DI}, where $\bf u,\bf v$ are the right and left principal eigenvectors of $A$. For these undirected networks, $\mbox{DI}_n=u_n^2$. After each node removal, components fragmented from the LSCC were also removed. 
In the upper panels of fig.~\ref{Attack}, we show the relative size of the LSCC after the removal of $k$ nodes (grey lines), whereas in the lower panels we show both $\lambda$ (dashed grey lines) and $\lambda_C$ (solid grey lines), where $C$ is given by eq. (\ref{C}). The horizontal dotted line denotes $\lambda_C=1$. For comparison, black lines indicate the the same variables for random node removal.
The disappearance of the LSCC corresponds to $\lambda_C\approx1$. For these undirected networks this corresponds to $\lambda-1\approx1$, whereas for directed networks we recover the result of ref. \cite{wPerc}, $\lambda\approx1$. Here $\lambda$ is the principal eigenvalue of the network adjacency matrix $A$.

\begin{figure}[t]
\onefigure[width=1\linewidth]{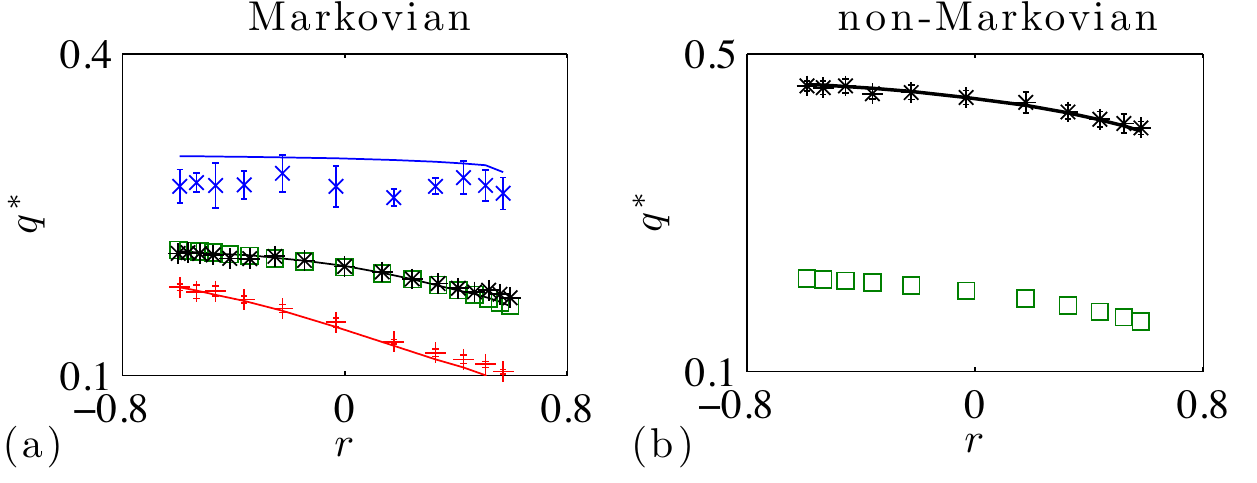}
\caption{(Colour online) 
(a) Values of $q^*$ predicted from eq.~(\ref{q1}) (solid lines) and observed experimentally (symbols) for percolation on an ER network having $N=10^4$ nodes and $5N$ undirected links rewired with correlations (measured by assortativity coefficient $r$). The three curves correspond to unweighted percolation ($l=0$, middle), and weighted percolation ($l=-1$, top, and $l=1$, bottom). Squares show the undirected Markovian prediction \cite{MarkUndir2}.
(b) Eq.~(\ref{q1}) (solid line) predicts observed values of $q^*$ (stars) for a non-Markovian network (see text), whereas the Markovian prediction (squares) cannot.}
\label{4}
\end{figure}

{In the remaining examples we demonstrate the robustness of our results to complex structures in networks. }Beginning with degree correlations, we first address Markovian correlations between the degrees of adjacent nodes, which can be characterized by the assortativity coefficient $r\in(-1,1)$ \cite{MarkUndir2}, where $r>0$ ($r<0$) indicates that nodes tend to connect to other nodes with similar (different) degrees. We again allow the retention probability of a node to depend on its degree, $q_n\propto d_n^{l}$.
We note that ref. \cite{bias} provides an ensemble approach for a similar removal strategy, but their analysis is restricted only to scale-free networks lacking correlations.
In fig.~\ref{4}a we show the effect of degree correlations on $q^*$ for an ER network with $N=10^4$ nodes and $5N$ undirected links that is rewired to have correlations. Experimental values are shown for $l=-1$ (x's), $l=0$ (stars), and $l=1$ (crosses). Solid lines indicate the prediction of eq.~(\ref{q1}), which was found to coincide with that of  eq.~(\ref{q2}), and the squares indicate the undirected Markovian prediction of ref.~\cite{MarkUndir2} (applicable only for $l=0$). Degree correlations were varied while keeping the degree distribution constant following the algorithm in refs.~\cite{MarkUndir2,DI}. 
Note that while assortativity promotes robustness for unweighted percolation ($l=0$) by reducing $q^*$, we find that its effect is reduced (amplified) for $l<0$ ($l>0$). For example, the percolation threshold is largely unaffected by degree corralations for $q_n\propto d_n^{-1}$ for this network (see x's in fig.~\ref{4}a).

Turning to degree-correlations of the non-Markovian type in undirected networks, a case which no previous theories can handle, we consider the set of networks used to produce fig.~\ref{4}a but subject them to the following rewiring process: each link $m \leftrightarrow n$ is replaced by two new links and a new node $j$, $m \leftrightarrow j \leftrightarrow n$, resulting in correlations across paths of length two. In fig.~\ref{4}b, whose horizontal axis is carried over from fig.~\ref{4}a, we show that the network-specific prediction eq.~(\ref{q1}) (solid line) agrees with the observed values of $q^*$ (stars) for unweighted percolation on these non-Markovian networks. For comparison, direct application of the Markovian ensemble method \cite{MarkUndir2} (squares) does not give good results. 

\begin{figure}[t] 
\onefigure[width=1\linewidth]{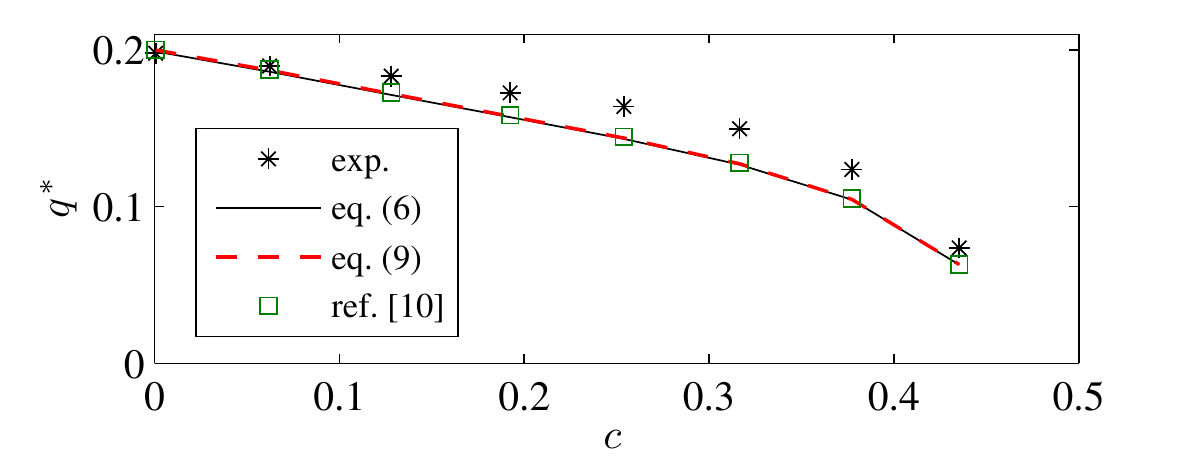}
\setlength{\belowcaptionskip}{-0.3cm}
\setlength{\abovecaptionskip}{-0.0cm}
\caption{{(Colour online) Predictions to $q^*$ and experimental values for percolation on an ER network with $N=10^4$ nodes and $5N$ links as a function of the clustering coefficient, $c$ (see text). Note that the lack of degree correlations yields good agreement between our results, eq. (6) and eq. (9), and the ensemble result of ref. \cite{Newman01}.}
\label{fig:clustering}}
\end{figure}

{We conclude by showing that our results remain accurate for relatively large clustering coefficient, $c$ \cite{clustering}. In this experiment links were iteratively added to an ER network by finding paths of length two and completing the triangles. In fig. \ref{fig:clustering} we show our predictions for and the observed value of $q^*$ as a function of $c$.}


\section{Conclusions}

We have presented a {\it network-specific} approach to weighted percolation in undirected networks and directed networks with bidirectional links (previously open problems). As opposed to most previous theory dealing with network ensembles, our method predicts unique percolation characteristics for each unique network. While ensemble and network-specific methods offer complementary strategies which may lead to different insights, in this Letter we highlighted several benefits of a network-specific approach. 
(i) Ensemble methods cannot account for variability across networks within the ensemble (see fig. \ref{11} and ref. \cite{ensemble}). 
(ii) Application of any ensemble approach to a real-world network requires {\it a priori} assumptions about that particular network ({\it e.g.,} that it either lacks complexity not accounted for in the ensemble or that its effects are small). 
(iii) Our approach naturally accounts for degree correlations of the non-Markovian type. 
%
%
(iv) Arbitrary targeting for node and/or link removal can be easily handled with our approach. 
(v) A network-specific analysis allows one to study deterministic attacks. Our results help explain why strategically decreasing and increasing $\lambda$ offer fundamental strategies to respectively attack \cite{DI} and protect \cite{dane} networks. 

{We have provided several examples showing that our results are robust to various attack strategies, degree distributions, degree correlations, and moderate clustering. However, when strong community structure is added we expect our analysis to breakdown as occurs with the ensemble approaches \cite{bagrow}. Finding a network-specific theory accounting for communities remains open for future research. 
Another direction of future work includes applying our techniques to more complicated percolation problems such as k-core percolation and Achlioptas processes. However, it is likely that efforts following the techniques provided here will also be restricted to the subcritical regime.}

\acknowledgments
We thank the NSF for financial support through Grant No. DMS-0908221.

\end{document}